\documentstyle[12pt]{article}
\def \F{\phi}

\def\NP{{\it Nucl. Phys.\ }}

\def\PL{{\it Phys. Lett.\ }}
\def\PR{{\it Phys. Rev.\ }}

\def\e{\epsilon}

\def\b{\beta}
\def\a{\alpha}

\def\s{\sigma}

\def\half{{1\over 2}}
\def\d{\dagger}
\def\be{\begin{equation}}
\def\eq{\end{equation}}
\def\Tr{{\rm Tr}}

\begin{document}

\begin{flushright}
OUTP-9404P\\
hep-th/9404058\\
\end{flushright}
\vspace{20mm}
\begin{center}
{\LARGE String Theory and Quantum Spin Chains}\\
\vspace{30mm}
{\bf Simon Dalley}\\
\vspace{5mm}
{\em Department of Physics\\
Theoretical Physics\\
1 Keble Road \\
Oxford OX1 3NP\\
England}\\
\end{center}
\vspace{30mm}

\abstract
The space-time light-cone hamiltonian $P^-$ of large-$N$
matrix models for dynamical
triangulations may be viewed as that of a quantum
spin chain and analysed in a mean field approximation.
As $N \to \infty$, the properties of the groundstate as a function of
the
bare worldsheet cosmological constant exhibit a parton
phase
and a critical string phase, separated by a transition with
non-trivial scaling at which $P^{-} \to -\infty$.

\newpage
\baselineskip .25in
\section{Introduction}

The quantization of non-critical relativistic string theories
remains a challenging and important problem, both from the point of
view of constructing superstring theories directly in four dimensions
and as a useful description of strongly-coupled gauge theories.
Non-critical strings differ from the critical ones in that physical
degrees of freedom are associated with the longitudinal oscillations.
Arguably the most powerful method for tackling this problem is the
 dynamical triangulation of the string worldsheet and subsequent
reformulation as a matrix field theory \cite{jan,rest}.
I.Klebanov and the author have suggested to capitalize on the well-known
simplicity
of string theories in light-cone formalism by performing light-cone
quantisation of the matrix models \cite{DKo}. A similar approach was described
many years earlier by C.Thorn \cite{CThorn}.
The light-cone quantisation has the advantage that  observables such as
the
spectrum and scattering amplitudes can be directly calculated. Some initial
numerical analysis of the 1+1-dimensional ($c=2$) matrix $\F^3$ theory in the
large-N limit was
carried
out in refs.\cite{DKo,kres}. In this paper a mean field theory will be
used to
determine the spectrum; these  results agree with the numerical ones
to
a certain extent, but throw up more structure which must be carefully
interpreted. In particular one finds two phases, a parton-like phase
and a critical-string phase, these being separated by a kind of
self-organising transition at which one can define a non-critical
string theory.
Although the 1+1-dimensional theory is studied principally, some of
the
conclusions will be seen to  carry over directly to higher dimensions.

\section{1+1-Dimensional $\F^3$ Theory}

In $c$ dimensions the field theory with action
\be
 S=\int d^c x \Tr \left( \half (\partial \F)^2 -\half \mu \F^2+
\lambda V(\F ) \right) \ ,\label{action}
\eq
where ${\F}_{ij}(x)$  is an $N$x$N$ hermitian matrix field, is used to
generate the $1/N$ planar diagram expansion \cite{Hoof}. If $V(\F )= \F^3
/3\sqrt{N}$, the dual graphs are triangulations and the theory is
finite after normal ordering in $c=2$ dimensions, the case which will
now
be discussed \cite{DKo}. Only the longitudinal oscillations of strings can
occur in this case. $\log{\lambda}$ is the bare worldsheet cosmological
constant
conjugate to area, and the Feynman propogator $\partial^2 + \mu$ is the
simplest
choice of link factor to specify the embedding in spacetime.
In light-cone quantisation, the fourier modes of $\F$ at fixed
light-cone
time $x^+$,
\be
\F_{ij}={1 \over \sqrt{2\pi}} \int_{0}^{\infty} {dk^{+} \over \sqrt{2k^{+}}}
(a_{ij}(k^{+}){\rm e}^{-ik^{+}x^{-}} + a_{ji}^{\d}(k^{+})
{\rm e}^{ik^{+}x^{-}})\ ,\label{Four}
\eq
satisfy standard creation and annihilation commutators;
\be
[a_{ij}(k^{+}),a_{lk}^{\d}(\tilde{k}^{+})] = \delta(k^{+} - \tilde{k}^{+})
\delta_{il}\delta_{jk}\ .\label{modeccr}
\eq
The normal-ordered light-cone energy $P^-$ and momentum $P^+$ are
\begin{eqnarray}
&:P^{-}: =
\half \mu \int_{0}^{\infty} {dk^{+}\over k^{+}} a_{ij}^{\d}(k^{+})
a_{ij}(k^{+}) \nonumber \\
&-{\lambda \over 4\sqrt{N\pi}}
\int_{0}^{\infty}
{dk_{1}^{+}dk_{2}^{+}\over \sqrt{k_{1}^{+}k_{2}^{+}(k_{1}^{+} + k_{2}^{+})}}
\left\{a_{ij}^{\d}(k_1^+ +k_2^+)a_{ik}(k_2^+)a_{kj}(k_1^+) +
a_{ik}^{\d}(k_1^+)a_{kj}^{\d}(k_2^+)a_{ij}(k_1^+ +k_{2}^+)\right\}
\label{nominus} \\
&:P^{+}: = \int_{0}^{\infty} dk^+ k^+ a_{ij}^{\d}(k^+ ) a_{ij}(k^+ )
\end{eqnarray}
where repeated indices are summed over and $\dagger$ does not act on
indices
but means quantum conjugate.

At a fixed $x^+ $ time-slice one considers the singlet states under
$\F \to {\Omega}^{\dagger} \F \Omega$, $\Omega$ unitary, as closed
strings;
the single string states are those elements of Fock space with all
indices contracted by one Trace,
\be
N^{-q/2} \Tr [a^{\d}(k_{1}^{+})\cdots a^{\d}(k_{q}^{+})] |0>\ ,\
\sum_{i=1}^{q} k_{i}^{+} = P^+\ .\label{string}
\eq
This represents a boundary of length $q$ in the dynamical
triangulation,
each $a^{\d}(k^+ )$ creating a parton of momentum $k^+$.
If one  lets $N \to \infty$ the multi-string states, corresponding to
more
than one Trace, can be neglected since
$1/N$ is the string coupling constant, and $P^-$ propagates strings
without splitting
or joining them. Also $(a^{\d}aa+ a^{\d}a^{\d}a)$ acts locally on
the string, coalescing two neighbouring partons in the Trace or
performing
the inverse process.
Since the states (\ref{string}) are already diagonal in $P^+ $, by
finding
the linear combinations which diagonalise also $P^- $ one determines
the
mass spectrum $M^2 = 2P^{+}P^- $ of the free $c=2$ string. Fixing
$P^+$, it is useful to discretise the momenta \cite{CThorn},
$k^+ = nP^{+}/K$,
for
positive integer\footnote{One excludes $n=0$ because $a^{\d}(0)$ modes
are presumably associated with non-perturbative vacuum structure,
whereas
here one is only interested in the Feynman diagrams obtained by
perturbation
theory about $\F =0$ as a fishnet approximation to string
worldsheets.}
  $n$ and some fixed large integer $K$ which plays the
role of cutoff. This renders the total number of states (\ref{string})
finite. Numerical analysis of the resulting eigenvalue problem was
done in refs.\cite{DKo,kres};
here an approximate analytic approach will be given
which allows the $K \to \infty$ continuum limit. The spectrum at
finite
$K$ is
given by
\be
{2P^{+}P^{-} \over \mu} = K (V -y T) \ ;\ y={\lambda \over 2\mu \sqrt{\pi}}
\ .\label{ham}
\eq
where $V$ and $T$ are discretised versions of terms appearing in
(\ref{nominus}) and $\mu$ sets the string tension scale. The
eigenfunctions
of $2P^{+}P^{-}$ give directly the structure functions of the string in
terms
of the Bjorken scaling variable $n/K$.

It is helpful to consider the states (\ref{string}) as possible spin
configurations on a $K$-site periodic chain, each site representing a
smallest fraction $1/K$ of the momentum of the string. The sites are
partitioned into $q\leq K$ partons by placing a down spin on any bond
between neighbouring sites belonging to different partons, while an up
spin on bonds between sites belonging to the same parton. The action
of
$P^-$ on this spin chain is in general very complicated, involving
interactions with a range that depends upon the state of the system
itself. In a mean field approximation however one may replace each
momentum variable occuring in $P^-$ (\ref{nominus}) by some typical
momentum per parton $K/\b$, $1\leq \b \leq K$, where $\b (y)$ is to be
determined self-consistently. In this approximation $V$ and $T$ become
single bond operators on the spin chain; namely,
\begin{eqnarray}
a^{\d} a & \equiv & \frac{1-{\s}_{3}}{2} \nonumber \\
a^{\d}a^{\d}a + a^{\d}aa & \equiv & {\s}_{+} + {\s}_{-}
\label{spins}
\end{eqnarray}
and one has the following 2x2 mean field hamiltonian at each
bond;
\[ \left( \begin{array}{cc}
0 & \frac{y\b^{3/2}}{\sqrt{K}} \\
\frac{y\b^{3/2}}{\sqrt{K}} & \b
\end{array}
\right) \]
 Thus the effect of the mean-field approximation is to replace
the true string structure function by a $\delta$-function at the
average parton momentum. This is a good approximation if the variance
is small such as when low momentum ``sea partons''
dominate. Strictly speaking, one should also project onto the spin states
invariant under translations along the chain  for exact equivalence
with
the states (\ref{string}), since Trace is cyclically invariant.
However
in the mean field approximation the translation operator is trivial
and
all its representations are degenerate.

The groundstate wavefunction (the ``tachyon''), which need not be
tachyonic in general, has one parton, therefore one spin down, at
$y=0$: $\Tr [a^{\d}(K)]|0>$. Diagonalising the mean-field hamiltonian
 one obtains a
consistency condition for $\b$, the average number of down spins;
\be
\b= \frac{2y^2 \b}{K +4y^2 \b - K\sqrt{1+ 4y^2 \b /K}}
+(K-1)\left( 1- \frac{2y^2 \b }{K+ 4y^2\b  -K\sqrt{1+ 4y^2 \b
/K}}\right)\ ,
\label{cons}
\eq
which for $K\to \infty$ and $\b$ finite becomes
\be
\b = \frac{1}{1-y^2}\ .\label{beta}
\eq
 The mass squared of this state is
\begin{eqnarray}
M^2 & = & \mu \b \left( \frac{K}{2} + \left( 1-\frac{K}{2} \right)
\sqrt{1 + 4y^2 \b /K} \right) \\
& =& \mu \b (1-y^2 \b )\ ;\ \b \ll K \nonumber \\
& =& \mu \left( \frac{1-2y^2}{(1-y^2 )^2}\right)\ .
\end{eqnarray}
Thus as $y$ is increased the average number of partons increases until
there is a critical point at $y=1$ where $\b \to \infty$. At this
point
$M^2 \to -\infty$. A comparison with the numerical result of ref.\cite{kres} is
made in fig.1. having adjusted the horizontal axis so that both curves
become tachyonic at the same point. (It is difficult to compare $y$
used
here with the analogous dimensionless variable
 employed in ref.\cite{kres} because one  could reasonably include
some coefficient of $O(1)$ in the off-diagonals of the mean-field
hamiltonian as a
refinement
of the approximation made.) The numerical results
are expected to be accurate when few partons dominate ($\b =2$ at
$M=0$) while the mean field argument should be better when there are
many partons. Thus the two methods give complementary results. In
particular one confirms the conjecture of ref.\cite{kres} that the
 critical point is characterised by $M^2 \to -\infty$. One further
confirms that the critical point signals a
transition
to a long wavelength regime for longitudinal string oscillations, since
$\b \to \infty$. Using (\ref{cons}) one can determine in more detail
how this happens. An ansatz $\b = \a K^{\gamma}$ at $y=1$ yields
$\gamma = 1/2$, $\a =1/\sqrt{3}$. For $y<1$, $\b$ is finite
(\ref{beta}),
while for
$y>1$ only $\gamma =1$ is consistent, with $\a$ satisfying
\be
\a^2 y^2 -2\a y^2 +(\a -1)(1-\sqrt{1+4\a y^2}) =0
\eq
for which $\a \to 1/2$ as $y \to \infty$. From this one
deduces that there is a non-trivial scaling law operating at the
transition point $y=1$, separating a parton phase with $\b$ finite
from a ``critical string'' phase in which the spin dynamics are
essentially those of the $\b =K$ critical string for which there are
no longitudinal oscillations i.e. it is analogous to the massive phase
of a conventional spin chain. The neighborhood of $y=1$ may allow one
to define a continuum string theory in non-critical spacetime
background. Indeed if one defines a renormalised worldsheet
cosmological constant through $y= 1-\e /\sqrt{K}$, then the
groundstate
satisfies $\b = \a \sqrt{K}$ with
\begin{eqnarray}
\a& = &\frac{-\e +\sqrt{\e^2 +3}}{3} \\
M^2 & = & -K\mu \a^2 + O(\sqrt{K})
\end{eqnarray}
This would imply that renormalised worldsheet area scales like string
length and that polymerised surfaces dominate the functional
integral as in Euclidean space \cite{jan}. The divergence of $M^2$ is similar
to that of critical string theory in light-cone gauge but it is not
clear at this stage how one should renormalise it to extract the
physically significant part. In critical string theory one can do it
by examining the $M^2$ of excited states.

The picture of the excited states coming from the mean field argument
is much less clear and does not agree well with the numerical solution.
Excited states are very badly represented even at $y=0$ since the
$q$-parton continuum, for example, is compressed into one infinitely
degenerate level at the threshold $M^2 =\mu q^2$. If the preceeding
analysis is formally repeated for the states having $q$ partons at
$y=0$ one gets
\be
\b = \frac{q}{1-y^2}\ ,\ M^2 = \frac{\mu q^2 (1-2y^2 )}{(1-y^2)^2}\ .
\eq
According to this, $y=1/\sqrt{2}$ is an inflexion point through
which the  spectrum for $q$ finite inverts itself
(giving continuous spectrum in the
neighbourhood). In particular the large $q$ states flip from
$+\infty$
to $-\infty$, which seems particularly unphysical. Comparison with the
numerical solution (fig.1.) is not good, yet the latter are
expected to be accurate at $y \sim 1\sqrt{2}$. Therefore this
tentatively suggests that the mean-field argument is not a very
reliable
guide
to the excitation spectrum.
Of course it could still be that the spectrum is continuous
at $y=1$, but at the moment this question, and the calculation of
string loop corrections, cannot be addressed without some reliable
scheme for the free-string excitation spectrum.

\section{$\F^4$ Theory}

Using a $\F^4$ theory should lead to similar results, being a
quadrangulation
rather than triangulation of the worldsheet. In this case an
important variational argument can be used, even in the presence of
some
transverse dimensions e.g. $c=4$. Using a transverse lattice action to
regulate the propogation in the transverse directions
\be
 S=\int d^2 x \Tr \left( \sum_{a}   \half (\partial \F_a)^2 -\half \mu
\F_{a}^{2}
+\frac{\lambda}{4N} \F_{a}^{4} + \sum_{<ab>} \F_a \F_b  \right)
\label{taction}
\eq
($<ab>$ indicates nearest neighbour sites $a$ and $b$ on the
transverse
lattice) the
light-cone
hamiltonian is
\begin{eqnarray}
:P^{-}:& =&
-\sum_{<ab>} \int_{0}^{\infty} {dk^{+}\over 2k^{+}} a_{a}^{\d}(k^{+})
a_{b}(k^{+}) + \sum_{a} \half \mu
\int_{0}^{\infty} {dk^{+}\over k^{+}} a_{a}^{\d}(k^{+}) a_{a}(k^{+})
\nonumber \\
& -&{\lambda \over 8N\pi} \int_{0}^{\infty}
{dk_{1}^{+}dk_{2}^{+}dk_{3}^{+}dk_{4}^{+}
\over \sqrt{k_{1}^{+}k_{2}^{+}k_{3}^{+}k_{4}^{+}}}
\left\{a_{a}^{\d}(k_1^+ )a_{a}^{\d}(k_2^+)a_{a}(k_3^+)a_{a}(k_4^+)
\right\} \delta (k_{1}^{+}+k_{2}^{+}-k_{3}^{+}-k_{4}^{+})\nonumber \\
& +&\left\{a_{a}^{\d}(k_1^+ )a_{a}(k_2^+)a_{a}(k_3^+)a_{a}(k_4^+) +
a_{a}^{\d}(k_2^+)a_{a}^{\d}(k_3^+)a_{a}^{\d}(k_4^+ )a_{a}(k_1^+
)\right\}
\delta (k_{1}^{+}-k_{2}^{+}-k_{3}^{+}-k_{4}^{+})  \nonumber \\
\end{eqnarray}
(The full index structure has been suppressed for clarity). This
 again acts locally on the string when $N \to \infty$, the
kinetic
terms taking three neighbouring partons in the Trace
into one, one into three, or
redistributing the momenta between two neighboring partons.
Discretising
$k^{+}$ and choosing as variational state
\be
|\Psi> = N^{-K/2} {\rm Tr}(a_{a}^{\d}(1)a_{a}^{\d}(1)\ldots
a_{a}^{\d}(1)) |0>\ ,
\eq
i.e. the maximum length string embedded at one particular site, the
light-cone hamiltonian satisfies
\be
\frac{<\Psi | P^{-} |\Psi >}{<\Psi | \Psi >} = \frac{K}{2P^+}\left( \mu
-{\lambda \over 4\pi} \right)\ .
\eq
Thus in the continuum limit $K \to \infty$ the groundstate mass
squared
tends to $-\infty$ for some $\lambda = \lambda_{c} < 4 \pi \mu$. This
further
confirms the existence and nature of the critical point found in the
previous section. It is natural to associate the phenomemena with the
divergence of planar graphs, which is expected to occur on general
grounds \cite{div}.

Returning to two dimensions, it would be nice to carry out a similar
mean field analysis of $\F^4$ theory. Unfortunately $P^-$ does not
factorise between bonds in this case. The best one can do is to
construct
an effective $\F^4$ theory by ``squaring'' $\F^3$ amplitudes, i.e.
decomposing the parton number-changing process $1 \to 3$ into
$1 \to 2$ then $2 \to 3$, which should capture the features for large
parton numbers.
This motivates the single bond hamiltonian
\[ \left( \begin{array}{cc}
0 & \frac{i\sqrt{x}\b^{3/2}}{\sqrt{K}} \\
\frac{-i\sqrt{x}\b^{3/2}}{\sqrt{K}} & \b (1-x)
\end{array}
\right) \]
where $x=\lambda /4 \pi \mu$. This form  has been chosen
because: $a^{\d}a^{\d}aa$ can move a down spin any distance along the
chain
up to $\sim K/\b$, so in mean field can be approximated by a single
bond
term $(K/\b ) {\rm x} (-x\b^2 /K)$ on the diagonal; $a^{\d}a^{\d}a^{\d}a$
creates two down spins, with amplitude $A^2$ say, where $A$ is the
amplitude to create one down spin, and since the second spin must be
created within distance $K/\b$ of the first there is a compensating
factor of $1/\b$ if $A$ creates a down spin at any bond, i.e.  $(A^2/\b )
=
(-x\b^2 /K)$. The groundstate analysis then proceeds as before, giving
\begin{eqnarray}
\b & =& \frac{1}{1-\frac{x}{(1-x)^2}} \\
M^2 & = & \frac{\mu (1-x)(1-\frac{2x}{(1-x)^2})}{(1
-\frac{x}{(1-x)^2})^2}
\end{eqnarray}
and the same critical behaviour. Beyond the critical point at
$x=(3-\sqrt{5})/2$ one finds $\a  \to 1$ at $x \to \infty$. Thus in the
strong coupling limit one recovers exactly the critical string \cite{CThorn}.

The mean-field hamiltonian  can also be considered at $x<0$,
corresponding
to the ``right-sign'' $\F^4$ theory.
A more appropriate parameterization in this case is to let $\lambda/4 \pi$
set the energy scale and $z=-4 \pi \mu / \lambda$ be the dimensionless
parameter;
\begin{eqnarray}
\b & =& \frac{1}{1-\frac{1}{(1+z)^2}} \\
M^2 & = & \frac{\lambda (1+z)^3 (z^2 +2z -1)}{(z^2 +2z)^2}
\end{eqnarray}
showing that tachyons appear for small enough parton mass squared $z$.
Indeed this appears to occur at $z>0$ indicating dynamical
symmetry breaking to a true groundstate which requires a more
careful treatment of zero modes to elucidate. The naive continuation
of the Fock vacuum $|0>$ into its tachyonic phase again takes one into
a critical string region eventually. Transverse lattice QCD at large
$N$
\cite{qcd} is very similar to the $\F^4$ theory, although the coupling
$\lambda$ can be both attractive  and repulsive, and it is natural to
expect a similar parton/string phase structure as the parton mass is
varied\footnote{Parton, i.e. transverse gluon, mass terms can appear on
renormalisation of the light-cone gauge QCD.}, as has been
suggested in the past by Klebanov and Susskind \cite{qcd}

\section{Conclusions}

The free non-critical bosonic string theory described by dynamical
triangulations has been shown  to possess a transition separating
a parton phase from a critical string phase. Introducing string
interactions, it would be interesting to see if the non-critical
string theory defined at the transition point is consistent in
any spacetime dimension. It is perhaps more appropriate, at least
eventually, to study the non-critical Green-Schwarz superstring
described by supertriangulations \cite{pari}. There appears to be no obstacle
to the numerical light-cone quantisation of this case, which always has
a tachyon-free true groundstate.
The spin-chain idea and its investigation by mean field, variational,
and exact solution,  has provided a partially
satisfactory analytic approach to complement the available numerical solutions
of light-cone quantisation of large-$N$ theories. In most cases one is
only interested in the lowest few energy levels and so use of
appropriate
algorithms could surely push the
attainable numerical results well beyond the present set. It is hoped
to address this in future work.

\vspace{10mm}
\noindent Acknowledgements: I would like to thank J.Cardy,
K.Demeterfi,
I.Klebanov, and A.Tsvelik for useful interactions.

\vspace{10mm}
\begin{center}
FIGURE CAPTION
\end{center}
\noindent Fig.1. -- A comparison of mean field results with the numerical
results of ref.[5] for the mass squared ($\mu =1)$. The solid
lines
are the mean field answer for the states corresponding to
one and two partons at $y=0$. The dotted lines are obtained by exact
solution at each $y$
for $K$ up to 15, then extrapolation to $K=\infty$ by fitting to a
ratio
of two polynomials in $K$. The numerical data contain the lowest two
states
of the two-parton continuum at $y=0$, which split at $y>0$ (they are always
degenerate in mean field). The variable $y$ used in this paper has
been
rescaled in this graph by $1/0.46\sqrt{2}$ so that the groundstates of
mean
field and numerical solution become massless at the same point.
\vfil
\end{document}